\documentclass[12pt]{article}
\RequirePackage[OT1]{fontenc}
\RequirePackage{amsthm,amsmath}
\RequirePackage[authoryear]{natbib}
\setcitestyle{aysep={}} 
\RequirePackage{graphicx}
\usepackage{pgf, tikz,float}
\usetikzlibrary{arrows,shapes,positioning}
\usetikzlibrary{calc,decorations.markings,automata}
\usepackage{multirow}
\usepackage{comment}
\RequirePackage[colorlinks,citecolor=blue,urlcolor=blue]{hyperref}
\usepackage{subcaption}

\newtheorem{assumption}{Assumption}
\newtheorem{theorem}{Theorem}

\newcommand\independent{\protect\mathpalette{\protect\independenT}{\perp}}
\def\independenT#1#2{\mathrel{\rlap{$#1#2$}\mkern2mu{#1#2}}} 
\newcommand{\bM}{\mathbf{M}}
\newcommand{\bY}{\mathbf{Y}}
\newcommand{\ACME}{\mbox{\tiny{ACME}}}
\newcommand{\ANDE}{\mbox{\tiny{ANDE}}}
\newcommand{\TE}{\mbox{\tiny{TE}}}
\newcommand{\RD}{\mbox{\tiny{RD}}}
\newcommand{\GEE}{\mbox{\tiny{GEE}}}
\newcommand{\one}{\mathbf{1}}
\newcommand{\newstuff}[1]{{\color{teal}#1}}

\renewcommand{\comment}[1]{{\bf [[#1]]}}

\newcommand{\blinding}[2]{#2}  

\begin{document}

\begin{center}

{\Large A Causal Mediation Model for Longitudinal Mediators and Survival Outcomes with an Application to Animal Behavior}

\medskip

Shuxi Zeng$^{1}$ \quad Elizabeth C.Lange$^2$ \quad Elizabeth A.Archie$^3$ \quad Fernando A.Campos $^4$ \quad \\ Susan C.Alberts $^{2,5}$ \quad Fan Li $^{1,\ast}$


\center{$^1$Department of Statistical Science, Duke University\\
$^2$Department of Biology, Duke University\\
$^3$Department of Biological Sciences, University of Notre Dame\\
$^4$Department of Antropology, University of Texas at San Antonio\\
$^5$Department of Evolutionary Anthropology, Duke University\\
$\ast$fl35@duke.edu, 214 Old Chemistry Building, Durham, NC, 27708, USA}

\end{center}
%
%
%
%
{\centerline{ABSTRACT}
In animal behavior studies, a common goal is to investigate the causal pathways between an exposure and outcome, and a mediator that lies in between. Causal mediation analysis provides a principled approach for such studies. Although many applications involve longitudinal data, the existing causal mediation models are not directly applicable to settings where the mediators are measured on irregular time grids. In this paper, we propose a causal mediation model that accommodates longitudinal mediators on arbitrary time grids and survival outcomes simultaneously. We take a functional data analysis perspective and view longitudinal mediators as realizations of underlying smooth stochastic processes. We define causal estimands of direct and indirect effects accordingly and provide corresponding identification assumptions. We employ a functional principal component analysis approach to estimate the mediator process, and propose a Cox hazard model for the survival outcome that flexibly adjusts the mediator process. We then derive a g-computation formula to express the causal estimands using the model coefficients. The proposed method is applied to a longitudinal data set from the Amboseli Baboon Research Project to investigate the causal relationships between early adversity, adult physiological stress responses, and survival among wild female baboons. We find that adversity experienced in early life has a significant direct effect on females’ life expectancy and survival probability, but find little evidence that these effects were mediated by markers of the stress response in adulthood. We further developed a sensitivity analysis method to assess the impact of potential violation to the key assumption of sequential ignorability.

\vspace*{0.3cm}
}
\clearpage

\section{Introduction\label{Sec_Intro}}

A common pursuit in biological studies is to understand mediation, that is, the causal relationships between an exposure or treatment $Z$, an outcome $Y$, and an intermediate variable (i.e. mediator) $M$ that lies on the causal path between $Z$ and $Y$. As a motivating example, consider an animal behavior study where we want to investigate the effect of early life adversity on survival outcomes, and how that effect is mediated through hormonal markers of the stress response in wild adult baboons. The classic mediation analysis method is the Baron-Kenny method, which fits two linear structural equation models (SEMs)---one on $Y$ predicted by $Z,M$ and one on $M$ predicted by $Z$---and interprets specific model coefficients as causal effects \citep{baron1986moderator,mackinnon2012introduction}. Recently there is a surge of research in combining the potential outcome framework for causal inference \citep{Neyman1923, Rubin1974} and the Baron-Kenny method \citep{robins1992identifiability,pearl2001direct, sobel2008identification, imai2010identification, tchetgen2012semiparametric,daniels2012bayesian, vanderweele2016mediation}. In particular, \cite{imai2010identification} proved that the Baron-Kenny estimator can be interpreted as a causal mediation estimator given a set of structural assumptions under the potential outcome framework. It has since led to many new methodological advancements and applications to disciplines beyond the traditional domains of SEM, including imaging, neuroscience, and environmental health \citep{lindquist2011graphical, lindquist2012functional,zigler2012estimating, kim2017framework, kim2019bayesian}. Advanced Bayesian modeling for mediation analysis has been also been developed \citep{daniels2012bayesian,kim2017framework,kim2018bayesian}. Comprehensive reviews on causal mediation analysis are given in \cite{vanderweele2015explanation} and in \cite{nguyen2019clarifying}.

Traditionally in mediation analysis the exposure $Z$, mediator $M$ and outcome $Y$ are all measured at a single time point. Recent studies increasingly involve time-varying data, where at least one of the triplet $(Z, M, Y)$ is measured repeatedly and the data pattern varies in specific applications. For example, in health studies, subjects' clinical information is often measured in multiple scheduled visits.  However, the majority of causal mediation research with time-varying data focuses on regularly observed data \citep{van2008direct,roth2012mediation,lin2017parametric, vanderweele2017timevarying}, and the analysis often utilizes marginal structural models \citep{robins2000msm}. Another line of research takes a functional data analysis perspective \citep{ramsay2005functional} when the observations are made on a dense grid. For example, motivated by applications in neuroimaging, \cite{lindquist2012functional} and \cite{zhao2018functional} view densely recorded functional magnetic resonance imaging (fMRI) mediators as functional data, and employed functional models as SEMs. 

None of the above methods is directly applicable to irregular longitudinal data. We can view sparse irregular longitudinal data from a missing data perspective, where each subject has an inherent value at any time point, but only a small number of these values are observed at irregular times. Therefore, any statistical analysis would in effect require imputing the unobserved values. A simple  method is to impute via ``last observation carried forward'' or the mean values between two consecutive observations. However, drawbacks of such n\"{a}ive methods are well documented in the literature on missing data \citep[e.g.][]{lachin2016fallacies, saha2009bias}. In particular, such methods implicitly assume that distribution of the missing data is exactly the same as that of the observed values, which is rarely true. Besides subject to large bias, these methods also fail to capture the dynamic nature of longitudinal data and tend to underestimate the uncertainty. Instead alternative model-based approaches are often more preferable. Specific to the context of modeling sparse irregular longitudinal data, the state-of-the-art method is the functional principal component analysis (FPCA) model \citep{yao2005functional,jiang2010covariatefpca,jiang2011functional,han2018functional, kowal2020bayesian}. This method treats the observed trajectories as realizations of underlying stochastic processes, and represents them as the sum of a few dominant functional principal components and consequently one can impute the entire process for any subject. The FPCA model  possesses desirable theoretical properties and has achieved much success in empirical applications. \cite{zeng2020causal} applied the FPCA method to causal mediation analysis with sparse longitudinal data. Specifically, they view the time-varying mediators and outcomes arising from a respective underlying stochastic process, and rigorously define corresponding causal estimands, and then imposed a FPCA model to the observed time-varying trajectories to estimate the estimands. In our specific application, such a functional model is natural from a scientific perspective because it is reasonable to contemplate animal behavior and physiology to follow an underlying smooth process and are observed with random errors. And thus to study the population pattern, it is key to impose a flexible model that can capture the dynamic feature of such smooth processes. 

\cite{zeng2020causal} focused on continuous outcomes, while survival outcomes are common in real applications, which is the focus of this paper. It is challenging to handle longitudinal mediators and survival outcomes simultaneously \citep{lange2012simple,vanderweele2011causal}. The existing related literature focuses on regular longitudinal mediators  \citep{zheng2017longitudinalsurvival,lin2017mediationsurvival}. There are two main complications in analyzing time-varying mediators with survival outcomes. First, the mediator value for a subject is not well-defined after that subject dies and thus the so-called cross-world counterfactual mediator values---key to causal mediation formulation---may sometimes be ill-defined. Second, the mediator is measured repeatedly over time, and prior survival is a prerequisite for later measurements and thus is a post-treatment confounder \citep{didelez2019defining,didelez2019causalmediationdefining, vansteelandt2019mediation}. \cite{didelez2019causalmediationdefining} 
tackled these problems by separating the primary treatment into a treatment on mediator and a treatment on outcome, and provided corresponding identification assumptions.

In this paper, we extend the functional data analysis method in \cite{zeng2020causal} to accommodate longitudinal mediators on an arbitrary grid with a survival outcome. Viewing the longitudinal mediator observations as functional data provides a principled and flexible way to adjust for the correlation between time-varying mediators and directly model its relationship with the survival outcomes, and also bypasses the two aforementioned complications (elaborated in Section \ref{Sec_Modeling}). We define relevant causal estimands and provide assumptions for nonparametrically identifying these estimands (Section \ref{Sec_Framework}). For estimation, we proceed under the two-SEM mediation framework \citep{imai2010identification}. Similar to \cite{zeng2020causal}, we specify a Bayesian functional principal component analysis (FPCA) model  \citep{kowal2020bayesian}  to project the mediator trajectories to a low-dimensional representation and impute the underlying mediator process. We also specify a Cox proportional hazard model for the survival outcome, and derive an analytical formula to express the causal estimands by the model coefficients using g-computation \cite{robins1986new} (Section \ref{Sec_Modeling}).   We apply the proposed method to a prospective and longitudinal observational data set from the Amboseli Baboon Research Project \citep{alberts2012amboseli} (Section \ref{Sec_Application}). We further developed a sensitivity analysis method to assess the impact of potential violations to the key assumption of sequential ignorability (Section \ref{Sec_Discussion}). 


\section{Motivating Application: Early Adversity, Physiological Stress, and Survival} \label{sec:background}
\subsection{Biological Background} 
Experiences during early life and adulthood can have profound effects on adult health and survival.  For example, negative socioenvironmental conditions during childhood are linked to dysregulation of the stress response and poor adult survival in humans \citep{RN3600, RN3488, RN3473, RN3564, RN3205, RN3474}.  In addition, dysregulation of the stress response in adulthood leading to altered glucocorticoid (GC) hormone profiles is hypothesized to reduce lifespan in humans \citep{RN3490, RN3604, RN3205, RN3518} and is known to do so in wild baboons \citep{RN3610}.
Can we identify the major mediators of early life adversity’s effects on adult survival?  On the one hand, the effects of early life adversity may be concentrated in one or several relatively simple health indices in adulthood, specifically dysregulation of the stress response \citep{RN3604, RN3205}. In this case, we would predict that GC hormone profiles are a major mediator of the link between early adversity and survival.  On the other hand, the effects of early adversity may be diffuse and multi-factorial, and/or variation in the adult stress response may have multiple causes, leading to very weak mediation by GC hormone profiles in the link between early adversity and survival.
No studies to date have been able to unambiguously link real time data on early life adversity, dysregulation of the stress response in adulthood (via assessment of adult GC profiles), and survival in the same individuals.  Therefore, the relative importance of early life adversity versus any independent effects of adult physiology in determining survival remains unclear \citep{RN3530, RN3200, RN3555}.

\subsection{Data}
\label{sec:data_general}
In this paper, we investigate the causal mediation relationship between early adversity, GC hormone profiles, and survival. We use data from a well-studied population of savannah baboons in the Amboseli ecosystem in Kenya. Founded in 1971, the Amboseli Baboon Research Project has prospective longitudinal data on early life experiences, and fine-grained longitudinal data on adult fecal GC concentrations \citep{alberts2012amboseli}. 

Our study sample includes 199 female baboons and 11914 observations in total. Survival was assessed for each female baboon starting at age 4 years, but GC hormone concentrations were measured only for females that had reached menarche (average age at menarche = 4.73 $\pm$ 0.56 years). For each subject, we had information on the experience of six sources of early adversity (i.e., exposure) \citep{tung2016cumulative, rosenbaum2020pnas}: drought, maternal death, close-in-age younger sibling, high group density, low maternal rank, and maternal social isolation. While only a small proportion of baboons experienced any given source of early adversity, most baboons experienced at least one source of early adversity. In our analysis we also create a binary exposure variable that indicates whether a baboon experienced any source of adversity.

\begin{figure}
\centering
\includegraphics[width=0.7\textwidth]{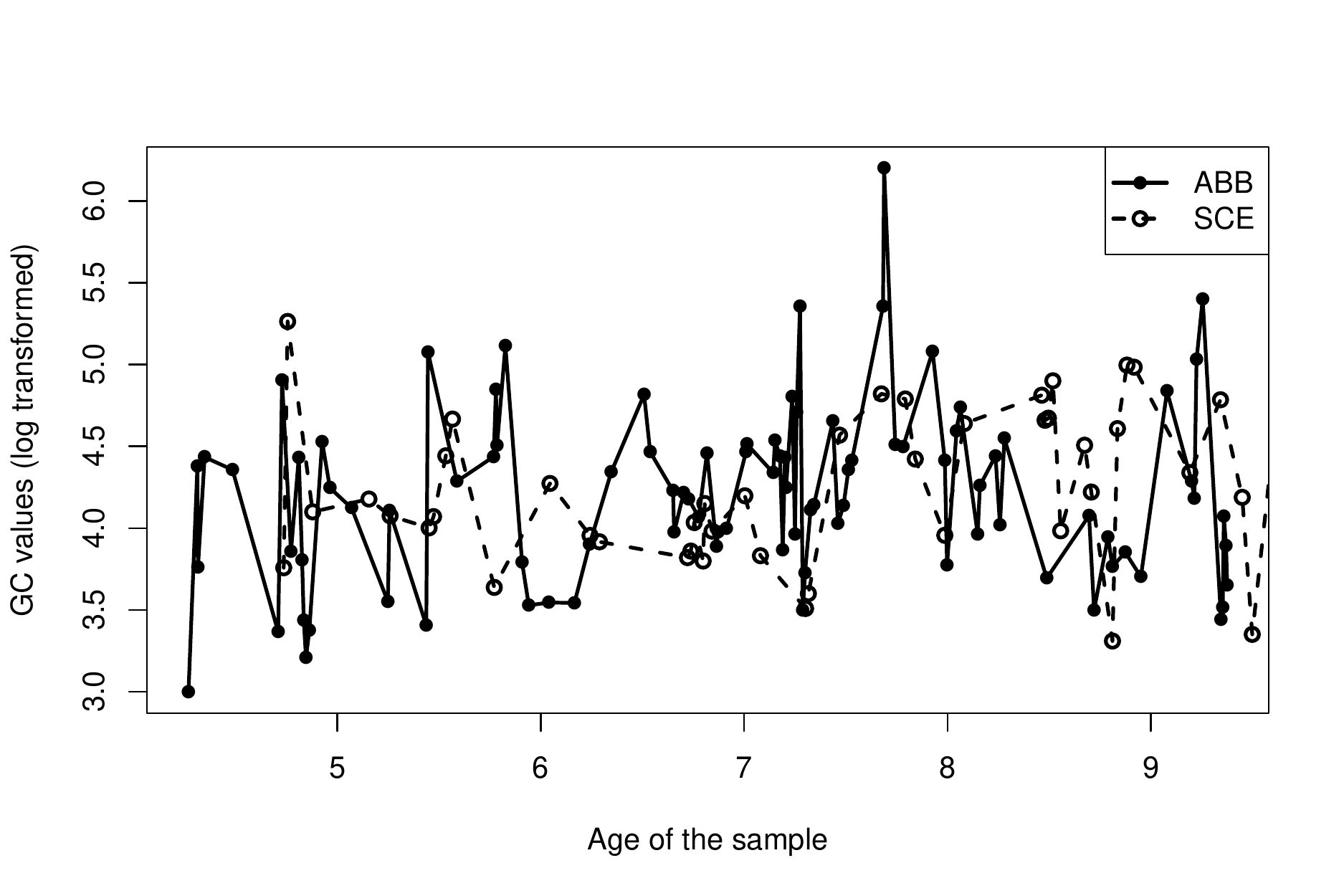}
\caption{Irregular and sparse GC observations (log transformed) for two randomly selected baboons in the sample.}
\label{fig:gc_trajectory}
\end{figure}

The mediator is each baboon's GC hormone profile across adulthood. These profiles are measured by assessing GC concentrations in fecal samples. For wild baboons, the GC hormone is recorded based on opportunistic collection of fecal samples and is thus measured on an irregular grid. The values of GC range from 7.51 ng/gm to 982.87 ng/gm with mean value at 76.90 ng/gm and standard deviation 39.58 ng/gm. We record the age of the subject at each sample collection as the time index for within-individual observations on GC concentrations. The frequency of observations and time grids of the mediator trajectories vary significantly between baboons: we have on average 59.86 GC observations of each baboon, but the number of observations of a single baboon ranges from 3 to 284.  Figure \ref{fig:gc_trajectory} shows the mediator trajectories of two randomly selected baboons (with codenames ``ABB'' and ``SCE'') in the sample. 

\begin{figure}
\centering
\includegraphics[width=0.5\textwidth]{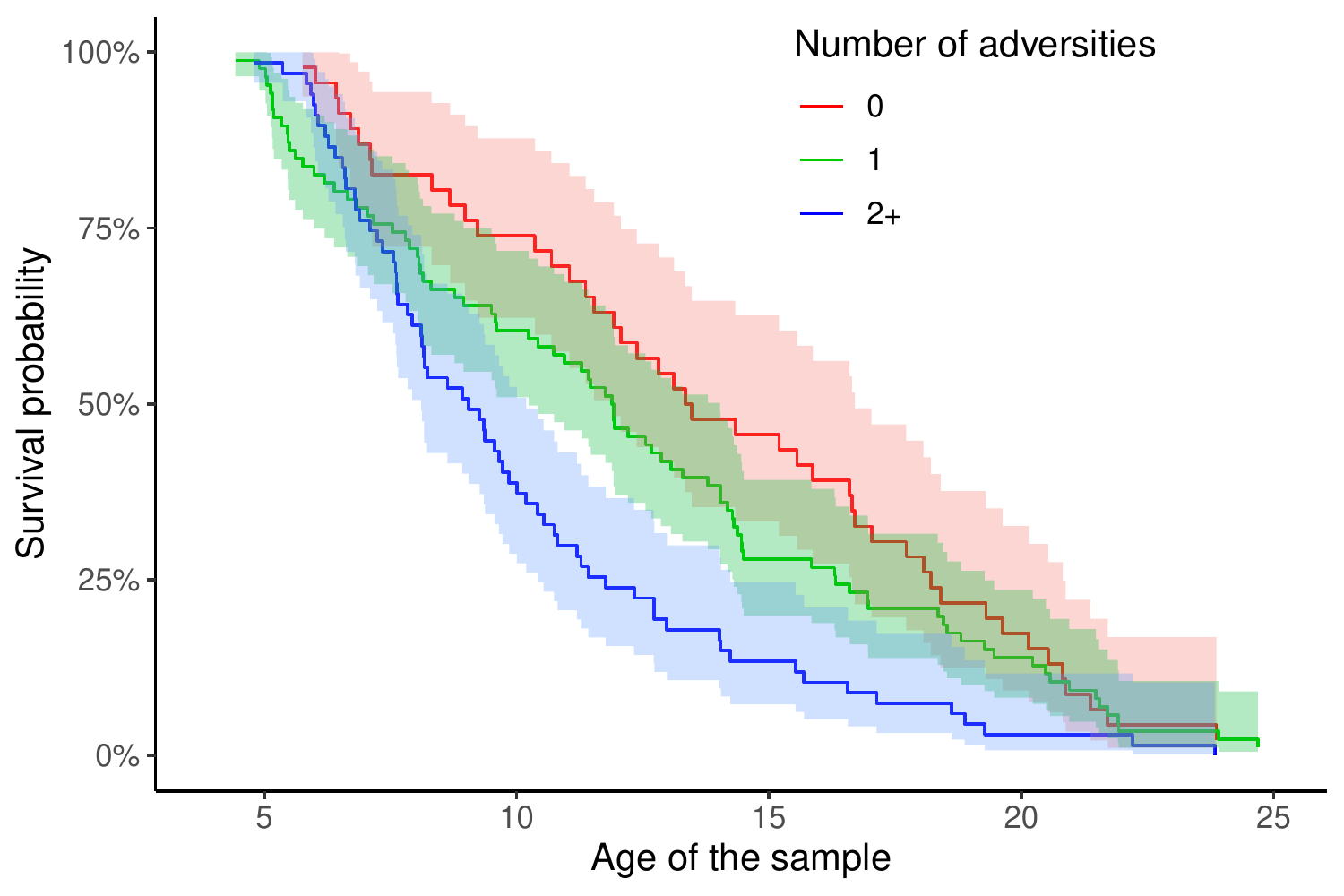}
\caption{Kaplan-Meier estimates of survival function in groups with different number of early adversities.}
\label{fig:km_group}
\end{figure}
The survival time is measured in years. Figure \ref{fig:km_group} shows the Kaplan-Meier estimates of the survival function in groups with different number of early adversities. Clearly the baboons experiencing fewer early adversities have better chance of survival. In particular, baboons who experienced two or more early adversities have a sharply decreased survival probability compared with those who had fewer adversities.

The time-varying covariates include reproductive state (i.e. cycling, pregnant, or lactating), density of the social group, group density squared \citep{RN3194}, max ambient temperature in the 30 days before the fecal sample was collected, whether the sample was collected in wet or dry season, the deviation in rainfall from expected during the three months prior to sample collection, storage time as fecal powder (time between collection of fecal sample and methanol extraction), storage time in methanol (time between methanol extraction and GC measurement), proportional dominance rank, and whether the focal female was top ranked or not. All these covariates are deemed important to wild baboons's physiology and behavior. More information can be found in \cite{rosenbaum2020pnas} and \cite{RN3426}.

\section{Causal Estimands and Identification} \label{Sec_Framework}
\subsection{Setup and Causal Estimands} \label{Sec_Setup}
Suppose we have a sample of $N$ subjects; each subject  $i\ (i=1,2,\cdots,N)$ is assigned to a treatment ($Z_{i}=1$) or a control ($Z_{i}=0$) group. For each subject $i$, we make observations at $T_{i}$ time points $\{t_{ij}\in [0,T], j=1,2,\cdots,T_{i}\}$, and the interval between two consecutive time points can differ within and across subjects. At each time point $t_{ij}$, we measure a mediator $M_{ij}$, and a vector of $p$ time-varying covariates
$\mathbf{X}_{ij}=(X_{ij,1},\cdots,X_{ij,p})'$. Let $V_{i}$ denote the survival time and $C_{i}$ be the censoring time. The survival time might be right censored when $C_{i}\leq V_{i}$ so we observe $\tilde{V}_{i}=\min(V_{i},C_{i})$ and the indicator that whether the subject failed within the study period $\delta_{i}=\textbf{1}_{V_{i}\leq C_{i}}$. In summary, we observe $(Z_{i},M_{ij},\mathbf{X}_{ij},\tilde{V_{i}},\delta_{i}),j=1,2,\cdots T_{i}$ for each subject $i$.

We view the observed mediator values drawn from a smooth underlying process $M_{i}(t)$, $t\in[0,T]$, with errors drawn from Normal distribution:
\begin{eqnarray}
M_{ij}&=&M_{i}(t_{ij})+\varepsilon_{ij}, \quad \varepsilon_{ij}\sim \mathcal{N}(0,\sigma_{m}^{2}).
\end{eqnarray}
We aim to investigate the relationship between $Z_{i}$, the stochastic processes $M_{i}(t)$, and the survival outcome $V_{i}$. In particular, we wish to answer two questions: (a) how big is the causal impact of the treatment on the survival time, and (b) how much of that impact is mediated through the mediator process? 

Following the standard notation of potential outcomes in causal inference \citep{ImbensRubin2015}, we move the time index of the mediator process to the superscript: $M_{i}(t)=M_{i}^{t}$ from now on. Also, we use bold font to denote a process until time $t$: $\mathbf{M}_{i}^{t}\equiv \{M_{i}^{s},s\leq t\}\in \mathcal{R}^{[0,t]}$. Similarly, we denote covariates between the $j$th and $j+1$th time point for subject $i$ as  $\mathbf{X}_{i}^{t}=\{X_{i1},X_{i2},\cdots,X_{ij'}\}$ for $t_{ij'}\leq t<t_{ij'+1}$. Further, let $\mathbf{M}_{i}^{t}(z)\in \mathcal{R}^{[0,t]}$ for $z=0,1$ and $t\in[0,T]$, be the potential values of the unobserved smooth mediator process for subject $i$ until time $t$ under the treatment status $z$; let $V_{i}(z,\mathbf{m})\in\mathcal{R}^{[0,T]}$ be the potential survival time for subject $i$ under the treatment status  $z$ and the mediator process taking value of $\mathbf{m} \in \mathcal{R}^{[0,T]}$. {In particular, $\mathbf{V}_{i}(z,\mathbf{M}_{i}^{t}(z'))$ is the potential survival time corresponding to exposure $z$ and the mediator value being the potential mediator under exposure $z'$. }
For each subject, we can only observe one realization from the potential mediator process and at most one potential survival time if not being censored: 
\begin{eqnarray}
&&\mathbf{M}_{i}^{t}=\mathbf{M}_{i}^{t}(Z_{i})=Z_{i}\bM_{i}^{t}(1)+(1-Z_{i})\bM_{i}^{t}(0),\\
&&V_{i} = V_{i}(Z_{i},\mathbf{M}_{i}^{T}(Z_{i}))=Z_{i} V_{i}(1,\mathbf{M}_{i}^{T}(1))+ (1-Z_{i})V_{i}(0,\mathbf{M}_{i}^{T}(0)).
\end{eqnarray}
We define the survival function for the potential survival time when a subject's treatment status is $z$ and the mediator process takes the value as if the subject was treated by $z'$, as $S_{z,z'}(t)$,
\begin{eqnarray}
S_{z,z'}(t)=\textup{Pr}(V_{i}(z,\mathbf{M}_{i}^{T}(z'))>t), \textup{for any $z,z'=0,1$}.
\end{eqnarray}

When $z\neq z'$, the potential outcome $\mathbf{V}_{i}(z,\mathbf{M}_{i}^{t}(z'))$ is called \emph{cross-world counterfactual} \citep{imai2010general} because the initial intervention (one world) is different from the hypothetical intervention for the mediator (another world). Cross-world counterfactuals are philosophically controversial \citep{lok2016defining,lok2021causal}; {they are particularly problematic in survival outcomes: a subject may survive longer in `one world’ than in the other `counterfactual world’ so that the mediator value is not well-defined in one of those two worlds.} 

We define the total effect (TE) of the treatment on the expected survival time as:
\begin{eqnarray}
\label{Definition_ATE_Process}
\tau_{\TE}^{h,t}&=&E[h\{V_{i}(1,\mathbf{M}_{i}^{T}(1));t\}-h\{V_{i}(0,\mathbf{M}_{i}^{T}(0));t\}].
\end{eqnarray}
where $t$ is a fixed time point, and $h(\cdot;t)$ is a function that transforms the survival outcome and thus defines causal estimands on different scales. For example, when $h(x;t)=x \wedge t$ (i.e. the truncation function), $\tau_{\TE}^{t}$ compares the restricted mean survival time. If we let $t\rightarrow \infty$, $\tau_{\TE}^{t}$ reduces to the standard average treatment effect (ATE) that compares the expected difference. When $h(x;t)=\textup{1}_{\{x>t\}}$ (i.e. the at-risk function), $\tau_{\TE}^{t}$ becomes the comparison on survival probability.
TE can be decomposed into direct and indirect effects \citep{robins1992identifiability, pearl2001direct,imai2010general}. Specifically, we define the average causal mediation (or indirect) effect (ACME) and the average natural direct effect (ANDE): for $z=0,1$
\begin{eqnarray}
\label{Definition_ACME_Process}
\tau_{\ACME}^{h,t}(z)&\equiv&E[h\{V_{i}(z,\mathbf{M}_{i}^{T}(1));t\}-h\{V_{i}(z,\mathbf{M}_{i}^{T}(0));t\}],\\
\tau_{\ANDE}^{h,t}(z)&\equiv&E[h\{V_{i}(1,\mathbf{M}_{i}^{T}(z));t\}-h\{V_{i}(0,\mathbf{M}_{i}^{T}(z));t\}].
\end{eqnarray}
ACME and ANDE quantifies the portion in the TE that goes through and bypasses the mediators, respectively. ACME is also referred as the \emph{natural indirect effect} \citep{pearl2001direct}, or the \emph{pure indirect effect} for $\tau_{\ACME}^{h,t}(0)$ and \emph{total indirect effect} for $\tau_{\ACME}^{h,t}(1)$ \citep{robins1992identifiability}.  It is easy to verify that TE is the sum of ACME and ANDE:
\begin{eqnarray}
\label{TE_decomposition}
\tau_{\TE}^{h,t}=\tau_{\ACME}^{h,t}(z)+\tau_{\ANDE}^{h,t}(1-z), \quad z=0,1.
\end{eqnarray}
Therefore, we only need to identify two of the three quantities $\tau_{\TE}^{h,t}$, $\tau_{\ACME}^{h,t}(z)$, $\tau_{\ANDE}^{h,t}(z)$. In this paper, we estimate $\tau_{\TE}^{h,t}$ and $\tau_{\ACME}^{h,t}(z)$, which can be expressed as functions of the survival function, $S_{z,z'}(t)$. Specifically, with the at-risk function $h(x;t)=\textup{1}_{\{x>t\}}$, we have
$$\tau_{\TE}^{h,t} = \int_{0}^{t} \{S_{1,1}(u)-S_{0,0}(u)\}\textup{d}u, \quad \tau_{\ACME}^{h,t}(z) = \int_{0}^{t} \{S_{z,1}(u)-S_{z,0}(u)\}\textup{d}u,$$
and with the truncation function $h(x;t)=x \wedge t$, we have
$$\tau_{\TE}^{h,t} = S_{1,1}(t)-S_{0,0}(t), \quad \tau_{\ACME}^{h,t}(z) = S_{z,1}(t)-S_{z,0}(t).$$
Further, for simplicity we only consider the estimands with $h=x \wedge t$ and $t=\infty$, which contrasts the expected potential survival time. Alternative estimands such as difference in restricted mean or survival probability \citep{vanderweele2011causal} can be derived in a similar manner within our framework.

\subsection{Identification assumptions}\label{Sec_Assumption}
Because we only observe a portion of all the potential outcomes, we need additional assumptions to identify causal estimands from the observed data. Below we present a set of assumptions that are sufficient for nonparametrically identifying ACME and ANDE .

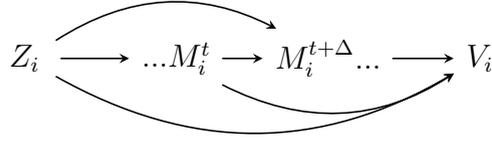
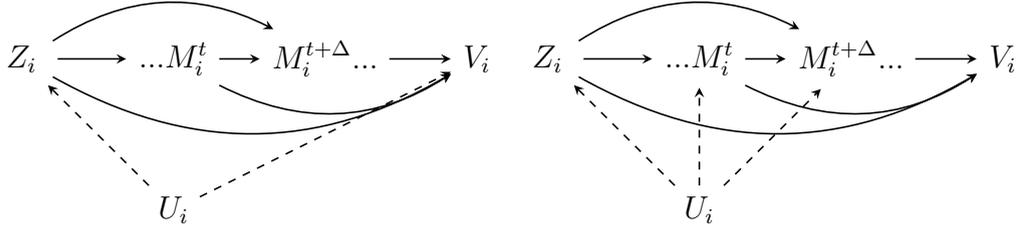
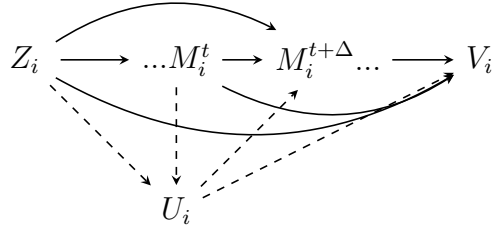
\begin{figure}[H]
\centering
\begin{subfigure}[b]{\textwidth}
\centering
\begin{tikzpicture}
[
> = stealth, 
shorten > = 0.5pt, 
auto,
node distance = 2cm, 
semithick 
]
\tikzstyle{every state}=[
draw = white,
thick,
fill = white,
minimum size = 3mm,
]

\node[state] (A)  (A){$Z_{i}$};
\node[right of= A] (B){...$M_{i}^{t}$};
\node[right of= B] (C) {$M_{i}^{t+\Delta}$...};%
\node[right of= C] (E) {$V_{i}$};%

\path[->] (A) edge node {} (B);
\path[->] (B) edge node {} (C);
\path[->] (C) edge node {} (E);
\path[->] (B) edge [bend right]  node  {} (E);
\path[->] (A) edge [bend right] node  {} (E);
\path[->] (A) edge  [bend left]  node {} (C);
\end{tikzpicture}
\caption{DAG of Assumption \ref{A.1} and \ref{A.2}}
\label{fig:DAG_A1A2}
\end{subfigure}

\begin{subfigure}[b]{\textwidth}
	\centering
	\begin{tikzpicture}
	[
	> = stealth, 
	shorten > = 0.5pt, 
	auto,
	node distance = 2cm, 
	semithick 
	]
	\tikzstyle{every state}=[
	draw = white,
	thick,
	fill = white,
	minimum size = 3mm,
	]
\node[state] (A)  (A){$Z_{i}$};
\node[right of= A] (B){...$M_{i}^{t}$};
\node[right of= B] (C) {$M_{i}^{t+\Delta}$...};%
\node[right of= C] (E) {$V_{i}$};%
\node[below of= B] (U) {$U_{i}$};%
\path[->] (A) edge node {} (B);
\path[->] (B) edge node {} (C);
\path[->] (C) edge node {} (E);
\path[->] (B) edge [bend right]  node  {} (E);
\path[->] (A) edge [bend right] node  {} (E);
\path[->] (A) edge  [bend left]  node {} (C);
\path[dashed,->] (U) edge  node {} (A);
\path[dashed,->] (U) edge  node {} (E);
\end{tikzpicture}
	\begin{tikzpicture}
	[
	> = stealth, 
	shorten > = 0.5pt, 
	auto,
	node distance = 2cm, 
	semithick 
	]
	\tikzstyle{every state}=[
	draw = white,
	thick,
	fill = white,
	minimum size = 3mm,
	]

\node[state] (A)  (A){$Z_{i}$};
\node[right of= A] (B){...$M_{i}^{t}$};
\node[right of= B] (C) {$M_{i}^{t+\Delta}$...};%
\node[right of= C] (E) {$V_{i}$};%
\node[below of= B] (U) {$U_{i}$};%
\path[->] (A) edge node {} (B);
\path[->] (B) edge node {} (C);
\path[->] (C) edge node {} (E);
\path[->] (B) edge [bend right]  node  {} (E);
\path[->] (A) edge [bend right] node  {} (E);
\path[->] (A) edge  [bend left]  node {} (C);
\path[dashed,->] (U) edge  node {} (A);
\path[dashed,->] (U) edge  node {} (B);
\path[dashed,->] (U) edge  node {} (C);
\end{tikzpicture}
\caption{DAG of two examples of violation to Assumption \ref{A.1} (ignorability)}
\label{fig:DAG_A1_violation}
\end{subfigure}
\begin{subfigure}[b]{\textwidth}
\centering
	\begin{tikzpicture}
	[
	> = stealth, 
	shorten > = 0.5pt, 
	auto,
	node distance = 2cm, 
	semithick 
	]
	\tikzstyle{every state}=[
	draw = white,
	thick,
	fill = white,
	minimum size = 3mm,
	]
\node[state] (A)  (A){$Z_{i}$};
\node[right of= A] (B){...$M_{i}^{t}$};
\node[right of= B] (C) {$M_{i}^{t+\Delta}$...};%
\node[right of= C] (E) {$V_{i}$};%
\node[below of= B] (U) {$U_{i}$};%
\path[->] (A) edge node {} (B);
\path[->] (B) edge node {} (C);
\path[->] (C) edge node {} (E);
\path[->] (B) edge [bend right]  node  {} (E);
\path[->] (A) edge [bend right] node  {} (E);
\path[->] (A) edge  [bend left]  node {} (C);
\path[dashed,->] (A) edge  node {} (U);
\path[dashed,->] (B) edge  node {} (U);
\path[dashed,->] (U) edge  node {} (C);
\path[dashed,->] (U) edge  node {} (E);
\end{tikzpicture}


	\caption{DAG of examples of violation to Assumption \ref{A.2} (sequential ignorability)}
	\label{fig:DAG_A2_violation}
\end{subfigure}
\caption{Directed acyclic graphs (DAG) of Assumptions \ref{A.1} and \ref{A.2}, and potential scenarios of violation, implicitly conditioning on the covariates $\mathbf{X}_{i}^{t}$ and a window between sufficiently close time points $t$ and $t+\Delta$ for \ref{A.2}. The arrows  represent a causal relationship, with solid and dashed lines standing for the measured and unmeasured relationships, respectively.  $U_i$ is an unmeasured confounder. The arrows between variables represent a causal relationship, with solid and dashed lines representing measured and unmeasured relationships, respectively. \label{Violation_DAG}}
\end{figure}

The first assumption extends the standard ignorability (or unconfoundedness) assumption and rules out the unmeasured treatment-outcome confounding. 
\begin{assumption}[Ignorability]
\label{A.1}
Conditional on the observed covariates, the treatment is unconfounded with respect to the potential mediator process and the potential survival time:
\begin{eqnarray*}
\{V_{i}(1,\mathbf{m}),V_{i}(0,\mathbf{m}),\mathbf{M}_{i}^{t}(1),\mathbf{M}_{i}^{t}(0) \}\independent Z_{i}\mid \mathbf{X}_{i}^{t},
\end{eqnarray*}
for any $t$ and $\mathbf{m}\in \mathcal{R}^{[0,t]}$.
\end{assumption}
In our application, Assumption \ref{A.1} indicates that there is no unmeasured confounding, conditioning on the observed covariates, between the early adversity, the process of adult physiological stress response, and survival. Equivalently, early adversity can be viewed as randomized among the baboons with similar covariates values. This assumption is likely to hold in our application because the early adversity events for the wild baboons were largely determined by nature. Assumption \ref{A.1} and two possible scenarios of violation are depicted by the directed acyclic graphs (DAG) in Figure \ref{fig:DAG_A1A2} and \ref{fig:DAG_A1_violation}, respectively.

The second assumption generalizes the sequential ignorability assumption in \citep{imai2010identification,forastiere2018principal} to the functional data setting.
\begin{assumption}[Sequential Ignorability]
\label{A.2}There exists $\varepsilon>0$, such that for any $0<\Delta<\varepsilon$, the increment of the mediator process from time $t$ to $t+\Delta$ is independent of the potential survival time conditional on the observed treatment status, covariates and the mediator process up to time $t$:
\begin{eqnarray*}
V_{i}(z,\mathbf{m}) \independent (M_{i}^{t+\Delta}-M_{i}^{t})\mid \{Z_{i},\mathbf{X}_{i}^{t},\mathbf{M}_{i}^{t}\},
\end{eqnarray*}
for any $z,0<\Delta<\varepsilon,t,t+\Delta\in [0,T],\mathbf{m}\in \mathcal{R}^{[0,T]}$.
\end{assumption}
In our application, Assumption \ref{A.2} implies that given the early adversity status, covariates, and the physiological stress history up to a given time point, change in the physiological stress within a sufficiently small time interval is independent of the potential survival outcome. 
Namely, we assume there are no unobserved mediator-outcome confounders in a sufficiently small time interval. Sequential ignorability and a scenario of its violation is depicted by the DAG in Figure \ref{fig:DAG_A1A2} and \ref{fig:DAG_A2_violation}, respectively. {One example of potential violation in the context of the Baboon study is genetic variation. Specifically, there may exist a genetic factor that affects both baboons' GC hormone level and survival, but we do not have genetic information of the baboons.} Overall, sequential ignorability is a strong assumption that may be violated in real world application. It is fundamental to causal mediation analysis, but is generally untestable even in randomized trials because it involves cross-world counterfactuals.  Therefore, a crucial part of causal mediation analysis is to conduct sensitivity analysis to assess the impact of potential violations to sequential ignorability, as we did in Section \ref{sec:SA}.

The third assumption imposes independent censoring mechanism, which allows us to identify the distribution of survival time from censored data.
\begin{assumption}[Independent censoring]
\label{A.3}The censoring time is independent of all remaining variables, including covariates, treatment, mediators and outcome:
\begin{eqnarray*}
C_{i}\independent \{X_{i},Z_{i},\textbf{M}_{i}^{t}(z),V_{i}(z,\mathbf{m})\},
\end{eqnarray*}
for any $z,t\in [0,T],\mathbf{m}\in \mathcal{R}^{[0,T]}$.
\end{assumption}
In our application, the time for a wild baboon to exit the study is largely random and therefore this assumption is deemed reasonable. {This assumption can be readily extended to a setting with covariate-dependent censoring, $C_{i}\independent \{\textbf{M}_{i}^{t}(z),V_{i}(z,\mathbf{m})\} \mid \{X_{i}, Z_{i}\}$. In that case, one would need to specify a model for the censoring mechanism and can combine analysis with the method of inverse probability of censoring weight to derive unbiased causal estimates \citep{robins2000correcting_IPC}.}

Under Assumptions \ref{A.1},\ref{A.2} and \ref{A.3}, we can identify the causal estimands, which is equivalent to identify the survival function $S_{z,z'}(t)$, nonparametrically from the observed data. This result is summarized in the following theorem.
\begin{theorem}
\label{T.1}
Let $t_{1}<t_{2}<,\cdots,<t_{k}<,\cdots,<t_{K}<\cdots$ be the time grid where we observe event ($\delta_{i}=1$) and consider a fixed time point $t$ that $t_{K}\leq t\leq t_{K+1}$. Under Assumptions \ref{A.1}, \ref{A.2} and \ref{A.3}, and some regularity conditions (specified in the Supplementary Material), the TE, ACME and ANDE can be identified nonparametrically from the observed data: for $z,z'=0,1$, we have
\begin{eqnarray*}
S_{z,z'}(t)&=& \int\int\textup{Pr}(V_{i}>t|\mathbf{M}_{i}^{t}=\mathbf{m},\mathbf{X}_{i}^{t}=\mathbf{x}^{t},Z_{i}=z)\textup{d}\textup{F}_{\mathbf{M}_{i}^{t}|Z_{i}=z',\mathbf{X}_{i}^{t}=\mathbf{x}^{t}}(\mathbf{m})\textup{dF}_{\mathbf{X}_{i}^{t}}(\mathbf{x}^{t})\\
&=&\int\int \{\prod_{k=1}^{K} \textup{Pr}(\tilde{V}_{i}>t_{k}|\tilde{V}_{i}>t_{k-1},\mathbf{M}_{i}^{t_{k}}=\mathbf{m}^{t_{k}},\mathbf{X}_{i}^{t_{k}}=\mathbf{x}^{t_{k}},Z_{i}=z)\}\times\\
&&\textup{d}\textup{F}_{\mathbf{M}_{i}^{t}|Z_{i}=z',\mathbf{X}_{i}^{t}=\mathbf{x}^{t}}(\mathbf{m})\textup{dF}_{\mathbf{X}_{i}^{t}}(\mathbf{x}^{t}),
\end{eqnarray*}
where $F_{W}(\cdot)$ and $F_{W|U}(\cdot)$ denotes the cumulative distribution of a random variable or a vector $W$ and the conditional distribution given another random variable or vector $U$, respectively.
\end{theorem}
We provide the proof of Theorem \ref{T.1} in the Supplementary Material. Theorem \ref{T.1} indicates that estimating the causal effects requires specifying two models: (a) the conditional survival probability given the treatment, covariates, and the observed mediator process,  $\textup{Pr}(V_{i}^{t}>t|Z_i,  \mathbf{X}_{i}^{t}, \mathbf{M}_{i}^{t})$, and (b) the conditional distribution of the observed mediator process given the treatment and covariates, $\textup{F}_{\mathbf{M}_{i}^{t}|Z_i, \mathbf{X}_{i}^{t}}(\cdot)$. These two models are in parallel to the two linear SEMs in the Baron-Kenny framework. In the next section, we specify these two  models and express the TE and ACME in terms of the model parameters.  

\section{Modeling mediators and survival outcome} \label{Sec_Modeling}
\subsection{Model for the mediators}
For the mediator process, we follow \cite{zeng2020causal} to employ a functional principal component analysis (FPCA) approach to impute the entire mediator process from sparse and irregular longitudinal data  \citep{yao2005functional,jiang2010covariatefpca,jiang2011functional}. In particular, we employ a Bayesian FPCA model similar to  \cite{kowal2020bayesian} to account for the uncertainty due to estimating the functional principal components \citep{goldsmith2013corrected}. The mediator model is the same as that in \cite{zeng2020causal} and thus we only present the main model form and refer the readers to \cite{zeng2020causal} for details. Our approach bypasses the two aforementioned conceptual challenges in causal mediation analysis with survival outcomes and time-varying mediators \cite{didelez2019causalmediationdefining} as follows. Conceptually, our setup here undertakes a type of cross-world counterfactual notion. Specifically, We conceive that every unit's mediator values exist in the entire span of time---regardless of its survival status in the actual world---in a counterfactual world. Therefore, the counterfactual mediator is defined even after a unit fails. Operationally, we use a functional model to impute all the counterfactual mediator values, but our estimation of causal effects implicitly conditions on unit's survival status, which ensures there is no causal effect of the current survival status on the mediators in the future. 

We assume the potential processes for mediators $\mathbf{M}_{i}^{t}(z)$ have the following Karhunen-Loeve decomposition,
\begin{gather}
\label{Mediator_Process}
M_{i}^{t}(z)=\mu_{M}(\mathbf{X}_{i}^{t})+\sum_{r=1}^{\infty}\zeta_{i,z}^{r}\psi_{r}(t),
\end{gather}
where $\mu_{M}(\cdot)$ are the mean functions of the mediator process $\mathbf{M}_{i}^{t}$; $\mathbf{\psi}_{r}(t)$ are the Normal orthogonal eigenfunctions for $\mathbf{M}_{i}^{t}$, and $\zeta_{i,z}^{r}$ are the corresponding principal scores of subject $i$. The above model assumes that the treatment affects the mediation processes only through the principal scores. We represent the mediator  process of each subject with its principal score $\zeta_{i,z}^{r}$. Given the principal scores, we can transform back to the smooth process with a linear combination. 

{The underlying process $\mathbf{M}_{i}^{t}(z)$ is not observed. Instead, we fit a truncated version of Model \eqref{Mediator_Process} to the observed mediator trajectories $M_{ij}$'s in group $z$. Specifically, we assume $M_{ij}$'s in the observed treatment group $Z_{i}$ to be randomly drawn from Model \eqref{Mediator_Process} truncated to its first $R$ principal components with errors:
\begin{gather}
\label{eq:mediator_model}
M_{ij} \mid Z_{i}=z \sim \mathcal{N}( X_{ij}'\beta_{M}+\sum_{r=1}^{R}\zeta_{i,z}^{r}\psi_{r}(t_{ij}),\sigma_{m}^{2}), \quad z=0,1,
\end{gather}
where $\psi_{r}(t)$ ($r=1,..., R$) are the orthonormal principal components, and $\zeta_{i,z}^{r}$ ($r=1,..., R$) are the corresponding principal scores for treatment $z$. Following the same parameterization in \cite{kowal2020bayesian}, we express the principal components $\psi_{r}(t)$ as a linear combination of spline basis. We also assume $\zeta_{i,z}^{r}$ follows a normal distribution with mean $\chi_{z}^{r}$ and diminishing variance $\lambda_{r}^{2}$ as $r$ increases: 
\begin{gather}	
\label{eq:mediator_pc_distribution}
\zeta_{i,z}^{r}\sim \mathcal{N}(\chi_{z}^{r}, \lambda_{r}^{2}), \quad \lambda_{1}^{2}\geq\lambda_{2}^{2}\geq\cdots\lambda_{R}^{2}\geq 0.
\end{gather}
We select the minimal truncation term $R$ which renders the fraction of explained variance (FEV), $\sum_{r=1}^{R}\lambda_{r}^{2}/\sum_{r=1}^{\infty}\lambda_{r}^{2}$ being greater than $90\%$. We usually require only 3 or 4 components to explain most of the variation. More details of the parameterization and specification of prior distributions can be found in the Supplementary Material.}

\subsection{Model for the survival outcome}
We posit the following Cox proportional hazards model for the survival time,
\begin{eqnarray}
\label{eq:survival_model}
\lambda(t_{ij}|X_{ij},Z_{i},\textbf{M}_{i}^{t})=\lambda_{0}(t_{ij})\exp\{\alpha Z_{i}+ X_{ij}'\beta_{S}+f(\textbf{M}_{i}^{t}; \gamma)\},
\end{eqnarray}
where $\lambda_{0}(t_{ij})$ is the baseline hazard rate,  and $f(\mathbf{M}_{i}^{t}; \gamma)$ is a function of the mediators with parameter $\gamma$, which captures the impact of the mediator process on the hazard rate. {Analysts can flexibly specify $f(\textbf{M}_{i}^{t}; \gamma)$ according to the specific application. For example, two common choices of $f$ are:}
\begin{enumerate}
    \item[(i)]  a \emph{concurrent model} that assumes the hazard rate depends on the instantaneous mediator value, $f(\textbf{M}_{i}^{t}; \gamma) = \gamma M_{i}(t)$; 
    \item[(ii)] a \emph{cumulative} model that assumes the hazard rate depends the entire mediator process until to time $t$, $f(\textbf{M}_{i}^{t}; \gamma) = \int_{0}^{t}\gamma(s) M_{i}(s)ds$.
\end{enumerate}

We can express the causal estimands, such as the TE and ACME, as functions of parameters of the mediator model \eqref{eq:mediator_model} and the survival outcome model \eqref{eq:survival_model}. First, we express $S_{z,z'}(t)$ via the g-formula as,
\begin{gather*}
S_{z,z'}(t)=\exp\{-\Lambda_{z,z'}(t)\},\\
\Lambda_{z,z'}(t) =\frac{1}{N}\sum_{i=1}^{N}\sum_{j=1}^{T_{i}} \lambda_{0}(t_{ij})\exp\{\alpha z+ X_{ij}'\beta_{S}+f(X_{ij}'\beta_{M}+\sum_{r=1}^{R}\chi_{z'}^{r}\psi_{r}(s); \gamma)\}(t_{ij}-t_{ij-1}),
\end{gather*}
where $\Lambda_{z,z'}(t)$ denotes the cumulative hazard for $V_{i}(z,\mathbf{M}_{i}^{t}(z'))$. Next, we can calculate $\tau_{\TE}$ and $\tau_{\ACME}$ based on $S_{z,z'}(t)$ with the equations in Theorem \ref{T.1}.

We impose a Gamma process prior for the baseline hazard rate $\lambda_{0}(t)$ \citep{fahrmeir2001bayesian,ibrahim2014bayeisan_survival,wang2013bayesian_survival} and standard normal prior distributions for other coefficients. For the cumulative model, we parameterize the function $\gamma(s)$ as a linear combination of the spline basis $\mathbf{b}(t)=(1,t,b_{1}(t),\cdots,b_{L}(t))'$  \citep{kowal2020bayesian}. Specifically,
\begin{align*}
\gamma(t)=\mathbf{b}(t)'\mathbf{p},
\end{align*}
where $\mathbf{p}$ is the coefficients with Normal prior, which enables a flexible modeling of how the past mediator history affects the survival outcome.

We perform posterior inference via Gibbs sampling. The credible intervals of the causal effects $\tau_{\TE}$ and $\tau_{\ACME}$ can be obtained from the posterior sample of the parameters in the model. We provide the details of the Gibbs sampler in the Supplementary Material.

\section{Application to the Amboseli Baboon Research\\ Project}\label{Sec_Application}

We apply the proposed method to the data described in Section \ref{sec:data_general} to investigate the causal relationship between early adversity, adult stress response, and survival in wild baboons. We perform a separate causal mediation analysis for each source of early adversity. We posit model \eqref{eq:mediator_model} for the GC concentrations, and Model \eqref{eq:survival_model} with the cumulative model of mediators for the survival outcome. {We have also fit the survival model with the concurrent model of mediators, results of which are similar to those of the cumulative model, and thus are omitted below.} In both models, we added two random effects, one for social group and one for hydrological year. In the mediator model, we use the log transformed GC concentrations instead of the original scale, which allows us to interpret the coefficient as the percent difference in GC concentrations between the adversity and non-adversity groups. 

\begin{figure}[ht]
\centering
\includegraphics[width=0.4\textwidth]{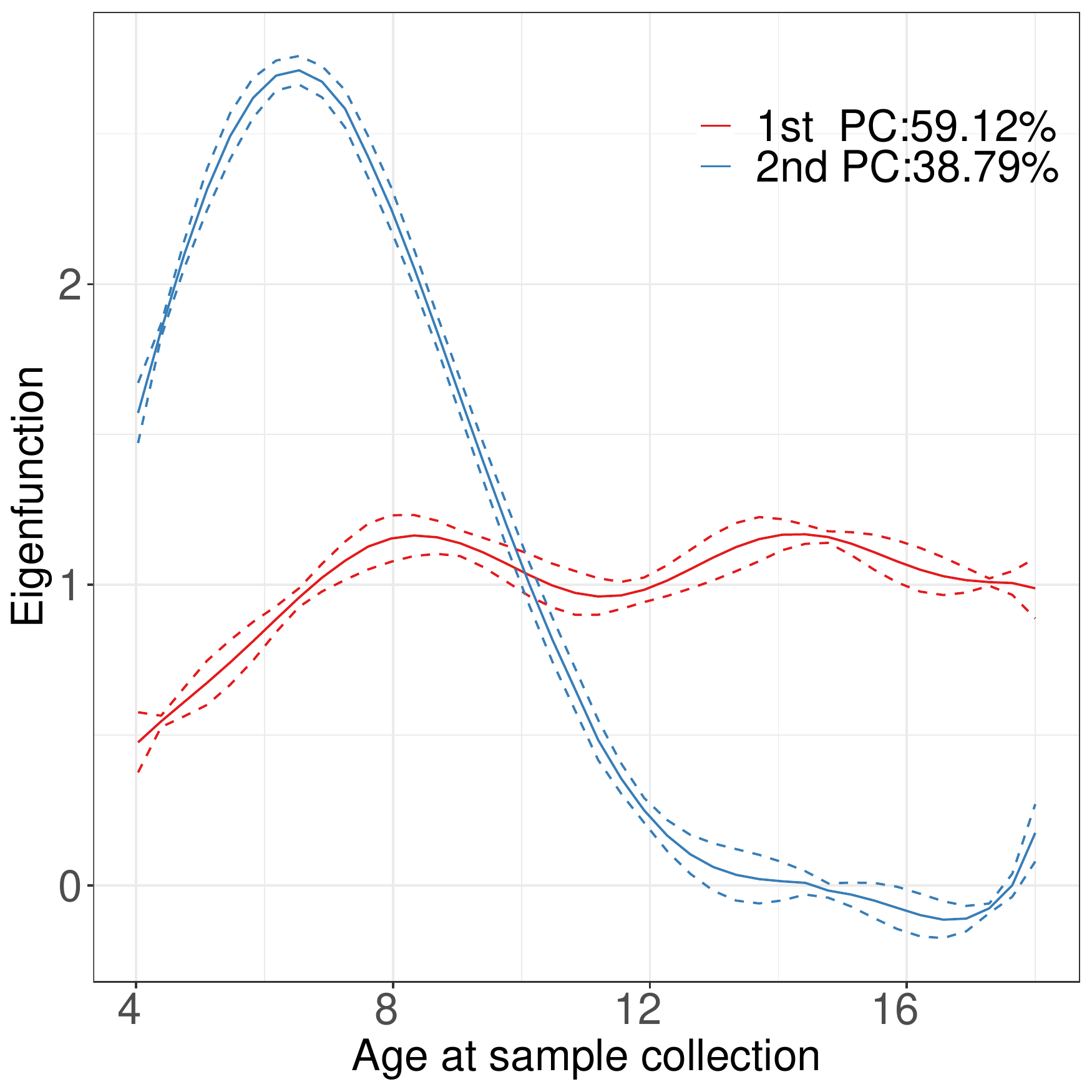}
\caption{The first two functional principal components of the mediator process, i.e., GC concentrations.}
\label{fig:pc}
\end{figure}

Here we first summarize the results of FPCA of the mediator trajectories, of which the first three functional principal components explain more than 90\% of the total variation. Figure \ref{fig:pc} shows the first two principal components extracted from the mediator process, which explain 59\% and 38\% of the total variation, respectively. The first component depicts a relatively stable trend throughout the life span. The second component shows a quick rise until age 6, then steady drop pattern across the lifespan.

\begin{figure}[ht]
\centering
\includegraphics[width=0.4\textwidth]{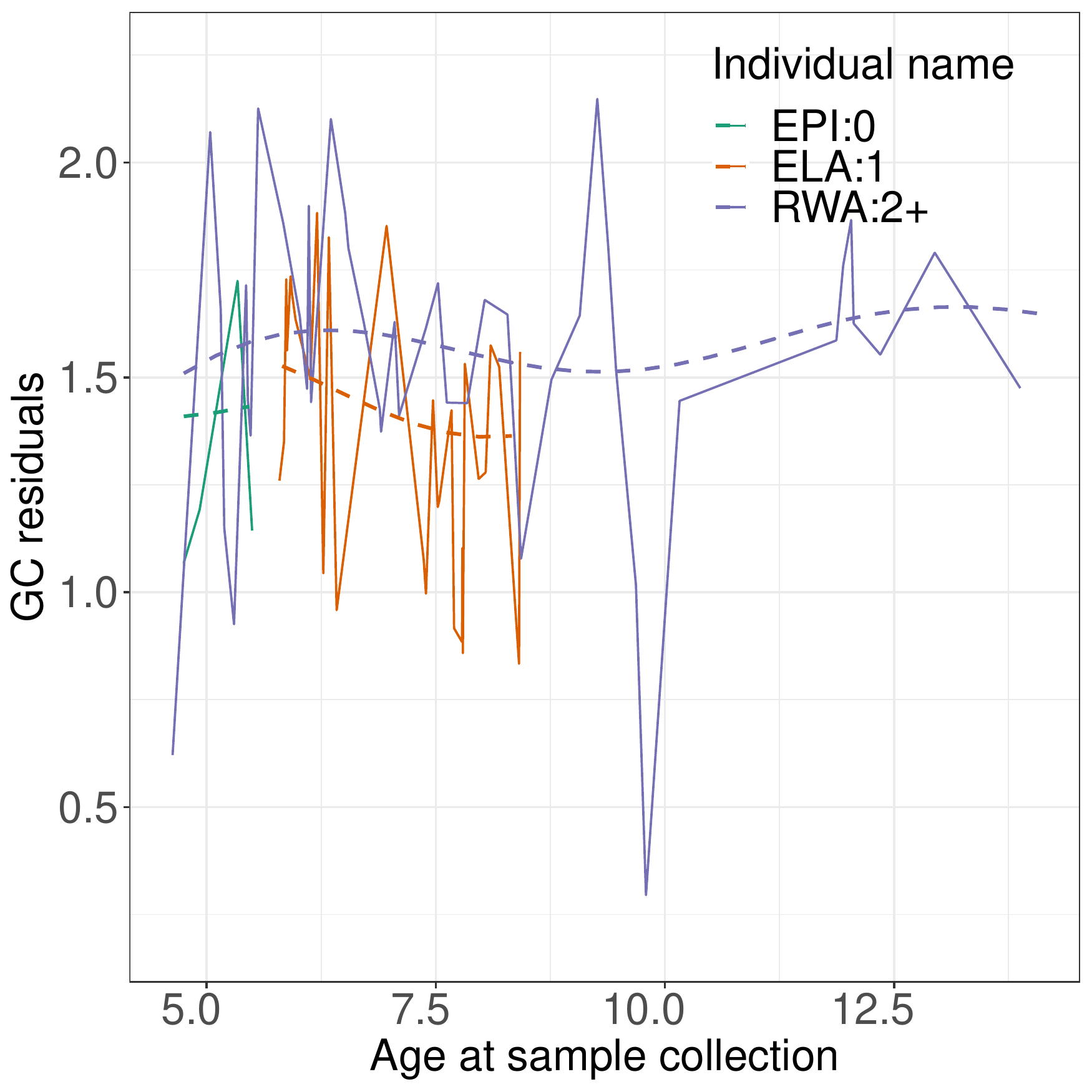}
\hspace{2em}
\includegraphics[width=0.4\textwidth]{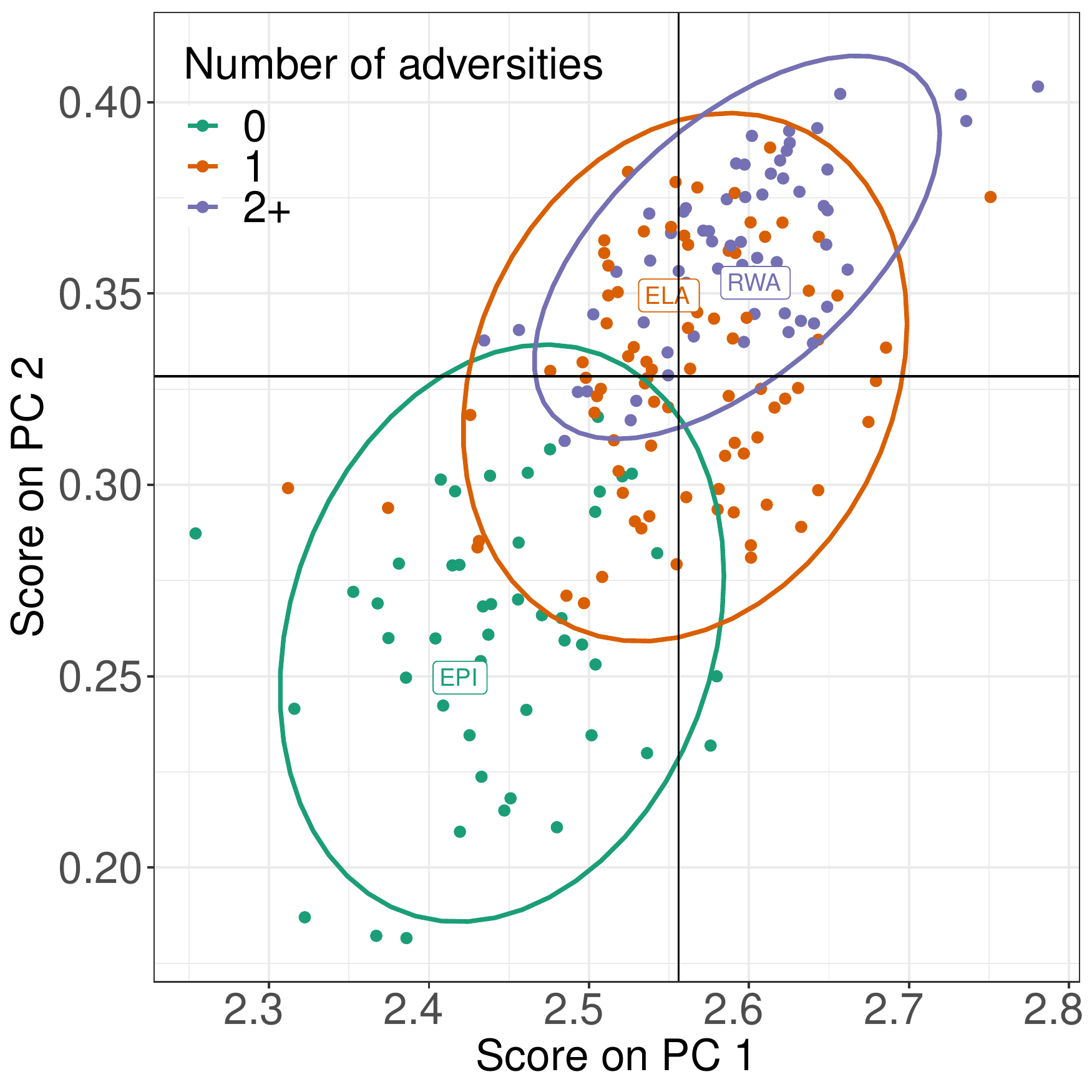}
\caption{Left panel: Examples of observed trajectory of GCs versus the posterior mean of its imputed smooth process of three baboons who experienced zero (EPI), one (ELA) and two (RWA) sources of early adversity, respectively. Right panel: Principal scores of the first (X-axis) versus second (Y-axis) principal component for the GC process of all subjects in the sample, including the three example subjects that are illustrated in the left panel (with individual names labeling the corresponding points). Color-coding is based on the number of early adversities experienced.}
\label{fig:ordination}
\end{figure}

The left panel of Figure \ref{fig:ordination} displays the observed trajectory of GCs versus the posterior mean of the imputed smooth process of three randomly chosen baboons who experienced zero (EPI), one (ELA), and two (RWA) sources of early adversity, respectively. We can see that the imputed smooth process generally captures the overall time trend of each subject while reducing the noise in the observations. Recall that each subject's observed trajectory is fully captured by its vector of principal scores, and thus the principal scores of the first few dominant principal components adequately represent the whole trajectory.  The right panel of Figure \ref{fig:ordination} shows the principal scores of the first (X-axis) versus second (Y-axis) principal component for the mediator process of all subjects in the sample, color-coded based on the number of early adversities experienced. We can see that significant differences exist in the distributions of the first two principal scores between the group who experienced no early adversity and the group experienced exactly one or the group with more than one sources of adversity.  

We now summarize the results on the causal estimates. Table \ref{tab:results} presents the posterior mean and 95\% credible interval of
the total effect (TE), direct effect (ANDE) and indirect effect  mediated through the GC hormone level (ACME) of each source of early adversity on life expectancy, as well as the effects of early adversity on the mediator. First, from the first column of Table \ref{tab:results} we can see that experiencing any source of early adversity would increase the GC concentrations in adulthood, which is detrimental to the health of the baboon. The effect is particularly severe for those who experienced drought, high group density, maternal death or low maternal rank in early life. For example, compared with the baboons who did not experience any early adversity, the baboons who experienced drought in the first year of life have $9.7\%$ increase in GC response. Overall, experiencing at least one source of early adversity corresponds to GC concentrations that are $9.4\%$ higher in adulthood.

Second, from the second column of Table \ref{tab:results} we can see a strong negative total effect of early adversity on the life expectancy of female baboons. Baboons who experienced at least one source of early adversity had a life expectancy approximately 1.5 years shorter than their peers who experienced no early adversity. The range of total effect sizes across all individual adversity sources varies from 0.691 to 2.199 years life reduction and the point estimates are consistently toward a shorter survival time, even for the early adversity sources for which the credible interval includes zero. Among the individual sources of adversity, females who were born during a drought or experienced maternal death experienced a particularly drastic drop in life expectancy, with effect sizes of 1.795 and 2.199 years respectively.

\begin{table}[!ht]
\centering
\caption{Total, direct and indirect causal effects of individual and cumulative sources of early adversity on life expectancy in adulthood in wild female baboons (measured in years).  95\% credible intervals are in the parenthesis.}
\resizebox{\textwidth}{!}{
\begin{tabular}{lcccccc}
\hline
Source of adversity &effect on mediator (\%) &effect of mediator on survival& $\tau_{\TE}$ & $\tau_{\ANDE}$ &$\tau_{\ACME}$   \\
\hline
Drought&9.7\%&-2.166&-1.795&-1.596&-0.199\\
&(1.5\%,18.0\%)&(-3.805,-0.528)&(-3.300,-0.291)&(-2.852,-0.341)&(-0.594,0.197)\\
Competing sibling&6.9\%&-2.126&-0.994&-0.886&-0.108\\
&(1.4\%,12.5\%)&(-3.539,-0.713)&(-4.038,2.049)&(-2.894,1.121)&(-0.210,-0.006)\\
High group density&11.9\%&-2.115&-0.691&-0.449&-0.242\\
&(2.8\%,21.0\%)&(-3.650,-0.581)&(-3.122,1.740)&(-2.512,1.614)&(-0.460,-0.024)\\
Maternal death&9.7\%&-2.616&-2.199&-1.972&-0.227\\
&(1.5\%,17.9\%)&(-4.311,-0.920)&(-3.856,-0.543)&(-3.527,-0.418)&(-0.466,0.013)\\
Maternal social isolation&8.0\%&-2.080&-0.692&-0.572&-0.120\\\
&(1.7\%,14.4\%)&(-3.454,-0.706)&(-3.188,1.805)&(-2.763,1.618)&(-0.255,0.016)\\
Low maternal rank&11.5\%&-2.075&-1.392&-1.046&-0.346\\
&(2.6\%,20.4\%)&(-3.609,-0.541)&(-3.991,1.207)&(-3.201,1.108)&(-0.728,0.036) \\
At least one &9.4\%&-2.212&-1.494&-1.292&-0.202\\
&(1.8\%,17.0\%)&(-3.689,-0.735)&(-2.748,-0.239)&(-2.264,-0.320)&(-0.551,0.147)\\
\hline
\end{tabular}
}
\label{tab:results}
\end{table}

Third, while female baboons who experienced harsh conditions in early life have a lower life expectancy, we found no strong evidence that these effects were mediated by GC hormone profiles. Specifically, the mediation effect $\tau_{\ACME}$ (the fifth column in Table \ref{tab:results}) is relatively small; the increase in adult GC concentrations accounted for a reduction in life expectancy of 0.202 years, when comparing the baboons who experienced at least one early adversity to those did not, with a credible interval including zero. In terms of individual early adversity sources, only two out of six individual adversity sources have a negative mediation effect with credible intervals not including zero, and both effects are quite small. 
{Moreover, we also find a significant negative effect from mediator on the survival, with effect size at about two years. Namely, one unit increase in the adult GC concentrations gives rise to two years decrease in the life expectancy of the baboons approximately. }On the other hand, the direct effects $\tau_{\ANDE}$ (the third column in Table \ref{tab:results}) are much larger than the mediation effects. When comparing the baboons with or without experiencing any source of early adversity, the direct effect of early adversity on life expectancy was 6.4 times stronger than the mediation effect running through adult physiological stress response. Specifically, for females who experienced at least one source of early adversity, the direct effect accounts of 1.292 years reduction in life expectancy while the mediation effect through GC accounts for only 0.202 years drop in the average survival time. 

\section{Sensitivity Analysis} \label{sec:SA}
The sequential ignorability assumption (Assumption \ref{A.2}) rules out unmeasured confounding between mediator and outcome, and is key to our analysis. But arguably sequential ignorability is often questionable in practice (see two aforementioned examples of potential violation), and it is generally untestable from observed data. So we develop a sensitivity analysis method to assess the impact of potential violation to sequential ignorability. Given the complex structure of mediation analysis, we adopt a model-based approach with the unmeasured confounders as the augmented variables,  along the lines in  \cite{imai2010general,huang2020sensitivity}. Specifically, we introduce an unmeasured confounder $U_{i}$ to characterize the correlation between the mediator process and the survival outcome that is not captured by covariates $X_{ij}$. Without loss of generality, we posit $U_{i}$ to be binary. We expanded the mediator model \eqref{eq:mediator_model} and the outcome model \eqref{eq:survival_model} to accommodate $U$ as follows:
\begin{eqnarray}
\label{eq:sensitivity_model_m}
M_{ij}=M_{i}(t_{ij})+\varepsilon_{ij}=\beta_{M}^{T} X_{ij}+\sum_{r=1}^{R}\psi_{r} (t)\{\tau_{0}^{r}(1-Z_{i})+\tau_{1}^{r}Z_{i}\}+\zeta_{M} U_{i}+\varepsilon_{ij},\\
\label{eq:sensitivity_model_s}
\lambda(t|X_{ij},Z_{i},\textbf{M}_{i}^{t})=\lambda_{0}(t)\exp\{\alpha Z_{i}+X_{ij}'\beta_{S}+f(\mathbf{M}_{i}^{t};\gamma)+\zeta_{S} U_{i}\},
\end{eqnarray}
where $\zeta_{M}$ and $\zeta_{S}$ are the pre-specified sensitivity parameters that measure the correlation between the unmeasured confounder and mediator process and the survival outcome, respectively. When sequential ignorability holds, there is no unmeasured confounder that simultaneously correlates with the mediator process and survival outcome, and thus $\zeta_{M}\zeta_{S}=0$. When both $\zeta_{M}$ and $\zeta_{S}$ are non-zero, sequential ignorability is violated. Therefore, we use $(\zeta_{M},\zeta_{S})$ as the sensitivity parameters to measured the degree of violation to Assumption \ref{A.2}.

Our sensitivity analysis consists of the following steps. First, we choose a grid of values of the sensitivity parameters $(\zeta_{M},\zeta_{S})$. For example, we choose $(\zeta_{M},\zeta_{S})\in \{0,0.1,0.2,0.5,1\}\times \{0,0.1,0.5,1\}$ in our application. Second, with each fixed pair of $(\zeta_{M},\zeta_{S})$, we fit the models \eqref{eq:sensitivity_model_m} and \eqref{eq:sensitivity_model_s}. Compared with the original models, \eqref{eq:mediator_model} and \eqref{eq:survival_model}, here we need to have an additional step of simulating the unmeasured confounder $U_{i}$ given the observed data  $(Z_{i},X_{ij},\delta_{i},\tilde{T}_{i})$, $(\zeta_{M},\zeta_{S})$ and the other model parameters. Next, we estimate the mediation effect $\tau^{\ACME}$ from the posterior sample following the same procedure in Section \ref{Sec_Modeling}. We repeat the above steps with the all possible combinations of $(\zeta_{M},\zeta_{S})$ on the pre-specified grid and examine how variable the estimates of $\tau^{\ACME}$ are to the values of $(\zeta_{M},\zeta_{S})$, which reflects how sensitive the causal estimates are to the violation of Assumption \ref{A.2}.

\begin{figure}[ht]
\centering
\includegraphics[width=0.6\textwidth]{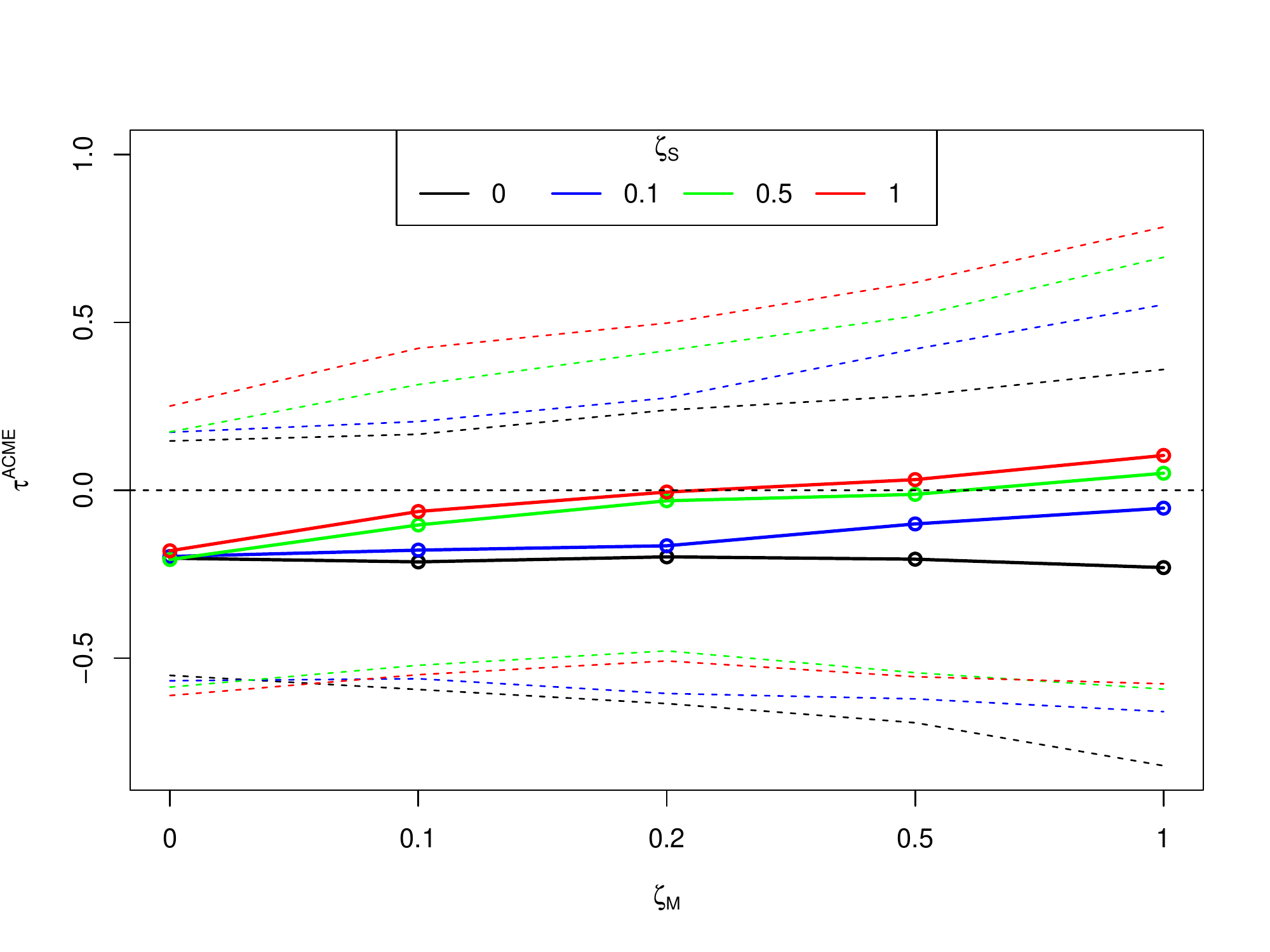}
\caption{Sensitivity analysis with a grid of $(\zeta_{S}, \zeta_{M})$. Each value of a fixed $\zeta_{S}$ is coded by a different color; for a given $\zeta_{S}$, the point estimate  and corresponding 95\% credible interval of $\tau^{\ACME}$ as a function  of $\zeta_{M}$ is presented by the solid and dashed line, respectively.} 
\label{fig:sensitivity}
\end{figure}

Figure \ref{fig:sensitivity} summarizes the results of the sensitivity analysis under the aforementioned specified grid of $(\zeta_{M},\zeta_{S})$ in our application. First of all, we notice that the point estimate of $\tau^{\ACME}$ becomes close to zero as $\zeta_{M}$ or $\zeta_{S}$ increases, and the effect size of $\tau^{\ACME}$ becomes negligible when $\zeta_{S}\geq 0.5$ and $\zeta_{M}\geq 0.1$. Also, the credible interval becomes wider when either one of the sensitivity parameters $(\zeta_{M},\zeta_{S})$ becomes larger. These patterns indicate our estimation of the mediation effect is sensitive to the sequential ignorability assumption. Recall that the our analysis under Assumption \ref{A.2} found only a small mediation effect. This sensitivity analysis further suggests that there is no strong evidence supporting that the adult physiological stress response mediates the effect between early adversity and survival.

\section{Discussion\label{Sec_Discussion}}
We proposed a method for causal mediation analysis with a longitudinal mediator on an arbitrary time grid and a survival outcome. The main idea is to view the time-varying mediator values as realizations from an underlying smooth process and use functional principal component analysis to impute the entire process, which is then used in the structural equation models. This approach naturally bypasses several conceptual and technical challenges in such settings. We defined several causal estimands in such settings and specified structural assumptions to nonparametrically identify these effects. We applied the proposed method to analyze the causal effects of early adversity on adult physiological stress responses and survival in wild female baboons. We found that experiencing adversity early in life significantly increases a baboon's GC response throughout its adulthood and decreases its survival probability. However, we found little evidence that the effect of early adversity on survival is mediated through the chronic elevation in the GCs, which is linked to poor health and survival in many species \citep{RN3518}. Our results suggest that early adversity and GC in adulthood have independent effects on survival and raise interesting questions in evolutionary biology about alternative causal pathways between early adversity and survival

We developed a model-based method to conduct sensitivity analysis regarding the key assumption of sequential ignorability. Given the complex structure of mediation analysis, related sensitivity analysis usually involves strong and sometimes overly simplified assumptions. For example, our sensitivity analysis depends on the correct specification of the mediator model and the outcome model, while misspecification is common in real applications. Also, to simplify the analysis we assume that the correlation structure between mediator and outcome is constant across time. Nevertheless, even a simplified sensitivity analysis still provides useful insights to causal mediation analysis; in particular it prevents over-interpreting the results and calls for more rigorous investigation of the causal assumptions. We notice that though sensitivity analysis has been standard in causal inference, it has not been routinely performed in causal mediation analysis. We believe more research on interpretable and flexible sensitivity analysis method would help the applied audience to employ causal mediation analysis. We acknowledge our conceptual setup, particularly the cross-world counterfactuals, might be controversial. But we argue various versions of such strong assumptions are generally required due to the complex nature of causal mediation problems with survival outcomes and longitudinal mediators. Our method offers an alternative approach that takes inferential advantage of a stochastic model of the longitudinal mediators, but conceptually it is not necessarily superior to the existing methods \citep{didelez2019causalmediationdefining}. In practice, researchers should always choose a causal inference method based on the plausibility of the key assumptions in their specific study. 

Though motivated by a specific application, the proposed method is readily applicable to other causal mediation studies with similar data structure. For example, comparative effectiveness studies increasingly use electronic health records (EHR) data, where the number of observations usually varies greatly between patients and the time grids are uneven. Moreover, many longitudinal studies in ecology rely on opportunistic sampling of their subjects, resulting in irregularly-spaced observations. 

The Supplementary Material can be found in \url{https://github.com/zengshx777/MFPCA_Codebase/blob/master/Mediation_FPCA_supp.pdf}. The data used in this paper is publicly available at the Duke Data Repository:
\url{https://research.repository.duke.edu/concern/datasets/bg257g02v?locale=en}.

\section*{Acknowledgements}
The majority of the data represented here was supported by the National Institutes of Health and the National Science Foundation, currently through NIH R01 AG053330 and R01 AG053308, as well as R01 HD088558, P01 AG031719, and NSF IOS 1456832. We also thank Duke University, Princeton University, and the University of Notre Dame for financial and logistical support. For assistance and cooperation in Kenya, we are grateful to the Kenya Wildlife Service (KWS), University of Nairobi, Institute of Primate Research (IPR), National Museums of Kenya, National Environment Management Authority, and National Commission for Science, Technology, and Innovation (NACOSTI). We also thank the members of the Amboseli-Longido pastoralist communities, and the Enduimet Wildlife Management Area for their cooperation and assistance in the field. Particular thanks go to the Amboseli Baboon Project long-term field team (R.S. Mututua, S. Sayialel, J.K. Warutere, Siodi, I.L.), and to T. Wango and V. Oudu for their untiring assistance in Nairobi. The baboon project database, Babase, is expertly managed by N. Learn and J. Gordon. Database design and programming are provided by K. Pinc.  This research was approved by the IACUC at Duke University, University of Notre Dame, and Princeton University and adhered to all the laws and guidelines of Kenya. For a complete set of acknowledgments of funding sources, logistical assistance, and data collection and management, please visit \url{http://amboselibaboons.nd.edu/acknowledgements/}.

\bibliography{FPCA_Reference}
\bibliographystyle{jasa3}

\end{document}